%% file: WSC-V2.tex
\documentclass{wscpaperproc}
\usepackage{latexsym}
\usepackage{graphicx}
\usepackage{mathptmx}
\usepackage{enumitem}
\usepackage[mathscr]{euscript}
\usepackage[ruled,noline]{algorithm2e}
\usepackage{subcaption}
\usepackage{amsmath}
\usepackage{amsfonts}
\usepackage{amssymb}
\usepackage{amsbsy}
\usepackage{amsthm}
\usepackage{bm}
\usepackage{xspace}
\usepackage{framed}

\usepackage[pdftex,colorlinks=true,urlcolor=black,citecolor=black,anchorcolor=black,linkcolor=black]{hyperref}
\hypersetup{nolinks=false}
\newtheorem{assumption}{Assumption}

\newtheorem{example}{Example}

\newtheorem{proposition}{Proposition}
\newtheorem{remark}{Remark}
\hypersetup{colorlinks = true, urlcolor = blue, linkcolor = black,
  citecolor = black }
\newcommand{\xmath}[1]{\ensuremath{#1}\xspace}

\newcommand{\RV}{\xmath{\mathscr{RV}}}
\newcommand{\MRV}{\xmath{\mathscr{MRV}}}
\newcommand{\R}{\xmath{\mathbb{R}}}
\newcommand{\mv}[1]{\bm{#1}}
\newcommand{\ttheta}{\mv{\theta}}

\newcommand{\qq}{\mv{q}}

\newcommand{\xx}{\mv{x}}
\newcommand{\XX}{\mv{X}}
\newcommand{\YY}{\mv{Y}}
\newcommand{\ZZ}{\mv{Z}}
\newcommand{\yy}{\mv{y}}

\newcommand{\pp}{\mv{p}}
\newcommand{\Varbeta}[1]{v_\beta{(\mv{#1})}}
\newcommand{\CVarbeta}[1]{C_\beta{(\mv{#1})}}
\newcommand{\e}{\mathrm{e}}
\newcommand{\Lev}{\xmath{\textnormal{lev}}}

\newcommand{\Real}{\mathbb{R}}

\newcommand{\Expc}{\mathbb{E}}

\newtheoremstyle{wsc}
{3pt}
{3pt}
{}
{}
{\bf}
{}
{.5em}
{}

\theoremstyle{wsc}
\newtheorem{theorem}{Theorem}

\begin{document}

%
%

\pagestyle{fancyplain}

\thispagestyle{plain}
\firstPageHead{}

\chead{\fancyplain{}{\itshape Deo, Murthy and Sarker}}

\rhead{}
\cfoot{}
\renewcommand{\headrulewidth}{0pt} 

\input{wscbib.tex}           

\setlength{\baselineskip}{12.7pt}

\title{COMBINING RETROSPECTIVE APPROXIMATION WITH IMPORTANCE SAMPLING FOR OPTIMISING CONDITIONAL VALUE AT RISK}

\author{Anand Deo\\
Karthyek Murthy\\
Tirtho Sarker\\[12pt]
	Singapore University of Technology and Design\\
    8 Somapah Rd, SINGAPORE 487372
}

\maketitle

\section*{ABSTRACT}
This paper investigates the use of  retrospective approximation solution paradigm in solving risk-averse optimization problems effectively via importance sampling (IS). While IS serves as a prominent means for tackling the large sample requirements in estimating tail risk measures such as Conditional Value at Risk (CVaR), its use in optimization problems driven by CVaR is complicated by the need to tailor the IS change of measure differently to different optimization iterates and the circularity which arises as a consequence. The proposed algorithm overcomes these challenges by employing a univariate IS transformation offering uniform variance reduction in a retrospective approximation procedure well-suited for tuning the  IS parameter choice. The resulting simulation based approximation scheme enjoys both the computational efficiency bestowed by retrospective approximation and logarithmically efficient variance reduction offered by importance sampling. 

\section{INTRODUCTION}
\label{sec:intro}
Conditional value at risk (CVaR) serves as a widely used risk measure towards assessing tail risks in  quantitative risk management and operations research (see \citeNP{mcneil2015quantitative,rockafellar2000optimization}). 
For a loss  $\ell(\XX,\ttheta)$ associated with a decision choice $\ttheta$ under a random realization $\XX,$ let  $\Varbeta{\ttheta}$ denote the $(1-\beta)$-th quantile of $\ell(\XX,\ttheta).$ Then its CVaR at the tail-level $\beta \in (0,1)$ is given by, 
\[
C_{\beta}(\ttheta) = E\left[\ell(\XX,\ttheta) \ \vert\ \ell(\XX,\ttheta) \geq \Varbeta{\ttheta}  \right], 
\]
which measures the average loss over the worst $\beta$-fraction of the realizations. Under the assumption that $\ell(\XX, \cdot)$ is  convex, $\CVarbeta{\ttheta}$ has been shown to possess favourable properties such as convexity, subadditivity and coherence (see \citeNP{acerbi2002coherence}). Minimizing CVaR $C_\beta(\ttheta)$ over a compact set $\Theta$ enjoys the following variational representation (see \citeNP{rockafellar2000optimization}), 
\begin{equation}\label{eqn:CVaR_OpT}
   c_\beta= \inf_{u\in \R,\ttheta\in \Theta} \left[u+ \beta^{-1} E\left(\ell(\XX,\ttheta)-u\right)^+  \right] =\inf_{u\in\R,\ttheta\in \Theta} f(u,\ttheta),
\end{equation}
which is a  convenient starting point for solving optimization problems. Consequently, CVaR has served as one of the primary vehicle for introducing risk aversion in a variety of planning and resource allocation problems in operations research and quantitative finance.

Since \eqref{eqn:CVaR_OpT} is rarely solvable in closed form, Monte Carlo methods are often deployed to obtain an approximate solution. Given $N$ i.i.d. samples of data $\XX_1,\ldots, \XX_N$ from the distribution of $\XX$, define the sample averaged objective,
\[
\hat{f}_{n}(u,\ttheta) = \left[u+\frac{1}{N\beta} \sum_{i=1}^N (\ell(\XX_i,\ttheta) - u)^{+}\right] 
\]
Then the sample average approximation (SAA) to the optimisation problem~\eqref{eqn:CVaR_OpT} may be constructed as
\begin{equation}\label{eqn:PMC}
\hat{c}_{n} = \inf_{u,\ttheta}  \hat{f}_n(u,\ttheta).
\end{equation}
Desirable large sample properties such as consistency and  asymptotic normality are well-known for SAA estimators (see \citeNP[Theorem 3.2]{shapiro1991asymptotic}) and their limiting variances scale inversely with the tail level $\beta$ of interest. This scaling is consistent with the understanding that one would need approximately $\tilde{O}(\beta^{-1})$ to witness loss scenarios exceeding the $(1-\beta)$-quantile $v_\beta(\ttheta),$ as $\beta \rightarrow 0.$ 
The computational effort in solving the resulting optimization problems becomes large as a consequence and it becomes necessary to resort to specialized sampling schemes such as importance sampling  (see, for example, \shortciteNP{he2021adaptive} and references therein) or aggregation sampling (\shortciteNP{Fairbrother_2019}).





In the evaluation of  objective $C_\beta(\ttheta)$ at a fixed decision choice $\ttheta,$ Importance Sampling (IS) is helpful if one can identify an alternate distribution under which the rare excess loss event $\{\ell(\XX,\mv{\theta}) \geq v_\beta(\ttheta)\}$ occurs more commonly. Indeed, IS has a rich literature on such change of measure prescriptions offering efficient variance reduction in the estimation of rare event probabilities (see, for example, \shortciteNP{Heidelberger,JUNEJA2006291}, and more recently, \shortciteNP{arief2021deep,deo2021achieving,ahn2021efficient}). For instance  with a good change of measure suited for tail probabilities of $\ell(\XX,\ttheta),$ \shortciteNP{glynn1996importance,glasserman2000variance} demonstrate how the variance reduction offered by the change of measure can be translated to efficient estimation of tail quantiles.  \shortciteNP{bardou2008computation,egloff2010quantile,he2021adaptive} develop adaptive algorithms which sequentially search for a good sampler choice within a IS distribution family in the estimation of CVaR.

In contrast to the prolific use of IS in tail estimation, its use in solving CVaR-driven optimization problems is limited by the availability of effective IS distribution families suited for complex objectives which arise in planning problems, and more severely, by the need to tailor the alternate sampling distribution choice differently for different decision choices: A change of measure which offers variance reduction for a decision choice $\ttheta \in \Theta$ may end up being inappropriate for a different decision choice $\ttheta^{\prime} \in \Theta$ (see Example \ref{eg:Example1} in Section \ref{sec:IS_algo} and Figure \ref{fig:Conc-Props} for an illustration). In a setting where identifying a good change of measure itself may require solving a non-trivial optimization problem (see, for eg.,  \shortciteNP{bai2020rare}), this dependence introduces a circular conceptual difficulty: A sampler choice which attains a small limiting variance depends, in turn, on the optimal decision choice which is the goal of our estimation to begin with. See \shortciteNP{he2021adaptive} for a detailed description of this circularity issue and how adaptive approaches in \shortciteNP{lemaire2010unconstrained,he2021adaptive} can be useful in overcoming this challenge.  The cross-entropy method by \shortciteNP{RUBINSTEIN199789}, which iteratively updates the IS distribution choice, by minimizing Kullback-Liebler divergence  between a proposed IS distribution and the theoretically optimal IS distributions for increasingly rare instances,  remains the most widely adopted approach.

The effectiveness of these adaptive schemes rely on working with a IS distribution family expressive enough to mirror the properties of theoretically optimal samplers at different decision choices $\ttheta \in \Theta.$ Their use, on the contrary, becomes practical if the chosen IS distribution family has a simple parametric representation (such as an exponential family) and is easy to sample from. Verifiably efficient prescriptions of IS distribution families have however remained elusive except in elementary instances involving piece-wise linear objectives and elliptical distributions modeling uncertainty in the problem. 

In this paper, we propose a simulation based approximation scheme which embeds importance sampling naturally in the well-known retrospective approximation solution paradigm for solving optimization problems (\shortciteNP{CHEN_2001,pasupathy2010choosing}). The retrospective approximation framework optimally balances the errors due to sampling and optimization approximations (see \citeNP{pasupathy2010choosing}) and lends itself naturally to the adaptive selection of IS parameter choices. 

For tackling the earlier highlighted challenges pertaining to the proposal and sampling of IS distribution family, we employ IS transformations which are structured sufficiently to induce distributions approximating the theoretically optimal zero variance measures. This deviates from the conventional approach of directly selecting a suitable IS distribution family, which has proven to be model and distribution specific and face scalability challenges. Specifically, we employ the single parameter family of  self-structuring transformations $\{\mv{T}_h(\XX):h>0\}$ proposed in  \shortciteNP{deo2021achieving}, where for  every $h>0,$ the mapping $\mv{T}_h:\R^d\to\R^d$ is a deterministic bijective function. 
Inheriting variance reduction properties exhibited in the context of tail probability estimation in \shortciteNP{deo2021achieving}, we exhibit uniform variance reductions offered by this IS transformation family in the evaluation of the objective $f(u,\ttheta)$ in \eqref{eqn:CVaR_OpT} over compact subsets of the decision set $\Theta.$ This ability to achieve  versatile variance reduction for a wide range of optimization iterates renders the resulting IS estimators as a natural choice towards untying the circularity issue highlighted earlier.  

Our proposal to use retrospective approximation together  with IS deviates from the existing adaptive IS literature in which  Robbins-Monro stochastic approximation remains the preferred solution paradigm. 
The  predominance of stochastic approximation in adaptive IS can be understood from the observation that samples from previous iterates are less suited to be used for the  gradient evaluation at the current iterate  due the differing changes of measure adopted in each iteration. Since our sampler is based on transformations of the underlying vector $\XX$ and enjoys robust variance reduction properties, it frees us to explore computationally attractive alternative solution paradigms. While our exposition treats the specific sampler family to be introduced shortly, we note that the enhanced retrospective approximation scheme developed in  Section~\ref{sec:SAA-IS-RA++} can be used in conjunction with other distribution families as well.

\noindent \textbf{Notation:} 
We use $\xrightarrow{{P}}$ to denote convergence in probability and $\Rightarrow$ to denote convergence in distribution. Boldface letters denote vectors. 
Likewise for a function $\mv{f}:\R^d\to\R^k$, $\mv{f}(\xx) = (f_1(\xx),\ldots, f_k(\xx))$.  We let $N(\mu,\sigma^2)$ denote a normal variable with mean $\mu$ and variance $\sigma^2$.  Let $\|\xx\|_p $ denote the $\ell_p$ norm of a vector $\xx\in \R^d$ and  
 $B_r(\mv{x})$ denote the $l_\infty$-metric ball of radius $r$ centred at $\xx$. For an increasing function $f:\Real \rightarrow \Real$, we let $f^{-1}$ denote its left-inverse.  For real valued functions $f$ and  $g$, we say that $f(x)=O(g(x))$ as $x\to\infty$ if there exist positive constants $M,x_0$ such that for all $x>x_0$, $|f(x)|\leq M|g(x)|$. We say that $f(x)=\tilde{O}(g(x))$ if $f(x)=O(g(x) \log^k(x)),$ for some $k>0$.  

\section{VARIANCE REDUCTION WITH IS TRANSFORMATIONS}\label{sec:IS_algo}
In this section we present an IS estimator for approximating CVaR objective. Recall our proposal to induce a suitable IS distribution via a transformation of the random vector $\XX$.  To accomplish this in our context, define the $\Real^d$-valued function $\mv{T}_h(\xx) := \xx [s_{h}]^{\mv{\kappa}(\xx)}$, where $s_{h} = h \log \log (1/\beta),$ with $h > 0,$ is a stretch factor which 
stretches the coordinates suitably via the exponent,
\begin{equation*}
\mv{\kappa}(\xx) := \frac{\log (1+|\xx|)}{\rho\|\log(1+|\xx|)\|_{\infty}},
\end{equation*}
in order to generate more samples in the extreme risk regions. The stretch factor $s_h,$ when viewed as a function of tail level $\beta,$ is increasing when the problem is made rarer by letting $\beta$ smaller (we henceforth drop the dependence on $\beta$ in our notation). Exponentiation is done component-wise in the above expression for $\mv{T}_h(\xx)$ as in, $\mv{T}_h(\xx) = (x_1 s_{h}^{\kappa_1(\xx)}, \ldots,x_d s_{h}^{\kappa_d(\xx)}).$ 
The map $\mv{T}_h:\Real^d \rightarrow \Real^d$ can be shown to be invertible almost everywhere on $\Real^d$ (see \citeNP[Proposition 1]{deo2021achieving}) and the resulting vector $\ZZ$ has a probability density if $\XX$ has a density. 
Letting $f_{\XX}$ and $f_{\ZZ}$ denote the respective densities of $\XX$  and $\ZZ,$ the likelihood ratio resulting from this change-of measure is given by, 
\begin{align}
\mathcal{L}_{h} = \frac{f_{\XX}(\ZZ)}{f_{\ZZ}(\ZZ)} = \frac{f_{\XX}(\ZZ)}{f_{\XX}(\XX)} J_h(\XX) 
\label{eq:LR}
\end{align}
An explicit expression of the Jacobian, $J_h(\xx) = \partial \mv{T}_h(\xx) /\partial \xx$ in \eqref{eq:LR}, given in \eqref{eqn:Jac} in the Appendix, can be obtained by replacing $(u/l)$ in \shortciteNP[Algorithm 1]{deo2021achieving} by $s_{h}.$ With this change-of-measure, we have the following unbiased estimator for the objective function in \eqref{eqn:CVaR_OpT}:
\begin{equation}\label{eqn:FIS}
\hat{f}_{is,n}(u,\ttheta) = \left[u+ \frac{1}{n\beta}\sum_{i=1}^n (\ell(\ZZ_{i},\ttheta) -u)^+\mathcal{L}_{h,i}\right],  
\end{equation}
where ${\XX}_1,\ldots,{\XX}_n$ are drawn i.i.d. from $\XX$, $\ZZ_{i}=\mv{T}_h(\XX_i)$ and $\mathcal{L}_{h,i}$ denotes the likelihood \eqref{eq:LR} evaluated at $\ZZ_{i}$.  Subsequently, one may optimise over the IS weighted objective \eqref{eqn:FIS}. The IS based estimator of minimum CVaR therefore becomes 
\begin{equation}\label{eqn:hatcis}
    \hat{c}_{is,n} = \inf_{u,\ttheta}\hat{f}_{is,n}(u,\ttheta)
\end{equation}
\subsection{A General Distribution Tail Model for Obtaining Variance Reduction Guarantees} We now outline the tail modelling framework under which the IS scheme described above provides substantial variance reduction.  We say that $f:\Real \rightarrow \Real$ is regularly varying if for all $x\in \R_+$,  $\lim_{t \rightarrow \infty} {f(tx)} /{f(t)} = x^{p}$, 
for some $p \in \Real$ (see \citeNP[Definition B.1.1]{de2007extreme}). In this case, we write $f \in \RV(p).$ We
say that a function $f:\R^d_+ \rightarrow \R_+$ is
\textit{multivariate regularly varying} if for any sequence $\xx_n$ of
$\Real^d_+$ satisfying $\xx_n \rightarrow \xx \neq \mv{0},$
\begin{align}
  \lim_{n \rightarrow \infty} n^{-1}f(\mv{h}(n)\mv{x}_n)
   = f^\ast(\mv{x}), 
  \label{MRV-het}
\end{align}
for some limiting $f^\ast: \R^d_{+} \rightarrow (0,\infty)$ and a
component-wise increasing $\mv{h}(t) = (h_1(t),\ldots,h_d(t))$
satisfying $h_i \in \RV(1/\rho_i), \ \rho_i > 0, i = 1,\ldots,d.$. 
\begin{assumption}\label{assume:Log-Weibull}\em
 The marginal distribution of $\XX = (X_1,\ldots,X_d)$ is such that the hazard functions of $\{X_i:i = 1\ldots, d$, $\{\Lambda_i: i  = 1.\ldots,d\}$ are eventually strictly increasing and $\Lambda_i \in \RV(\alpha_i)$ for some $\alpha_i > 0$. Further the
  density of $\XX$ when written in the form,
  \begin{align}
  f_{\mv{X}}(\mv{x}) = \exp(-\psi(\mv{x})) \ \xx \in \R_d^+, \text { satisfies $\psi \in \textnormal{\MRV}(\psi^\ast,\mv{h}).$}
    \label{X-pdf-form}
  \end{align}
  
\end{assumption}
A wide variety of parametric and nonparametric multivariate distributions, including normal, exponential family, elliptical, log-concave distributions and Archimedian copula models satisfy Assumption~\ref{assume:Log-Weibull}. Marginal distributions which satisfy $\Lambda_i \in \RV(\alpha_i)$ include all distributions that are either Weibull-type heavy-tailed or possess lighter tails (such as exponential, normal, etc.). 
\begin{assumption}\label{assume:L}
There exists a limiting function $\ell^*(\cdot; \cdot)$ and $\rho>0$, such that whenever $\xx_n\to \xx\neq \mv{0}$ and $\mv{\theta}_n\to\mv{\theta}$, 
\begin{equation}\label{eqn:MVconv}
n^{-\rho} \ell(n\xx_n; \mv{\theta}_n) \to \ell^*(\xx;\mv{\theta}).
\end{equation}
such that for every $\ttheta$, $\ell^*(\cdot,\ttheta)$ is a homogeneous function of $\xx$.
\end{assumption}
Assumption~\ref{assume:L} is satisfied for example, when $\ell(\xx,\ttheta) = c(\ttheta^\intercal \xx)$, such that for $x\neq 0$, $c(tx)/t^\rho \to c^*(x) $ as $t\to\infty$. It is also satisfied in more complicated examples, such as two-stage linear optimisation problems where
    \begin{equation}\label{eq:eg-two-stage}
        \ell(\xx,\ttheta) = \mv{c}^\intercal\xx + \inf_{\mv{A} \yy + \mv{b} =\xx} \ttheta^\intercal\yy.
    \end{equation}
We refer the reader to \citeNP[Section 2 and Appendix B]{deo2021achieving}  for a more elaborate discussion on the conditions under which Assumptions~\ref{assume:Log-Weibull} and \ref{assume:L} are satisfied.

\subsection{Uniform Variance Reduction Guarantees}  
One way to quantify the benefits gained by the use of IS transforms is to analyse the variance of the functions $\hat{f}_{1}$ and $\hat{f}_{is,1}$ as defined in \eqref{eqn:PMC} and \eqref{eqn:FIS}. Note that these are simply the sample variances of the objective functions in SAA and IS-SAA respectively.  Proposition~\ref{prop:IS-Uniform} below quantifies these.
\begin{proposition}\label{prop:IS-Uniform}\em
Suppose Assumption~\ref{assume:Log-Weibull} holds. Define the set $S_{r}=\{(u,\ttheta): u^*(1-r) \leq u  \leq u^*(1+r), \  \theta_i^*(1-r) \leq \theta_i  \leq \theta^*_i(1+r)\}$. Then for any $r,h_{\min},h_{\max} \in (0,\infty)$,
\begin{equation}\label{eqn:IS-Uniform}
    \lim_{\beta\to 0}\sup_{(u,\ttheta)\in S_{r}, h\in (h_{\min},h_{\max})} \left\vert\frac{\log \text{var}\left[[(\ell(\ZZ;\ttheta) -u)^+] \mathcal{L}_{h}\right]}{\log \text{var}([(\ell(\XX,\ttheta ) - u)^+])} -2\right\vert = 0 \text{ and }
\end{equation}

\begin{equation}\label{eqn:IS-Uniform-2}
    \sup_{(u,\ttheta)\in S_{r}, h\in (h_{\min},h_{\max})} \frac{\text{var}(\hat{f}_{is,1}(u,\ttheta)) }{\text{var}(\hat{f}_{1}(u,\ttheta))} = o\left(\beta^{\frac{1}{1+\rho(r)}}\right),
\end{equation}
where $\rho(r)$ decreases to $0$ as $r\to 0$.
\end{proposition}
Proposition~\ref{prop:IS-Uniform} quantifies the scale of variance reduction in the objective due to the IS transformation by comparing it with the sample average objective. The first part shows that the  asymptotic variance reduction is optimal when viewed in the logarithmic scale (see \shortciteNP{BJZWSC}). Further, since $r$ appears multiplicatively in $S_r$, equations \eqref{eqn:IS-Uniform} implies that a uniform reduction in the variance of the objective function is obtained over a substantial neighbourhood of the optimal point. On the other hand, \eqref{eqn:IS-Uniform-2} re-expresses this improvement in performance in terms of the rarity parameter $\beta$.

\begin{figure}[t]
      \begin{center}
      \begin{subfigure}{0.42\textwidth}
          \includegraphics[width=0.98\textwidth]{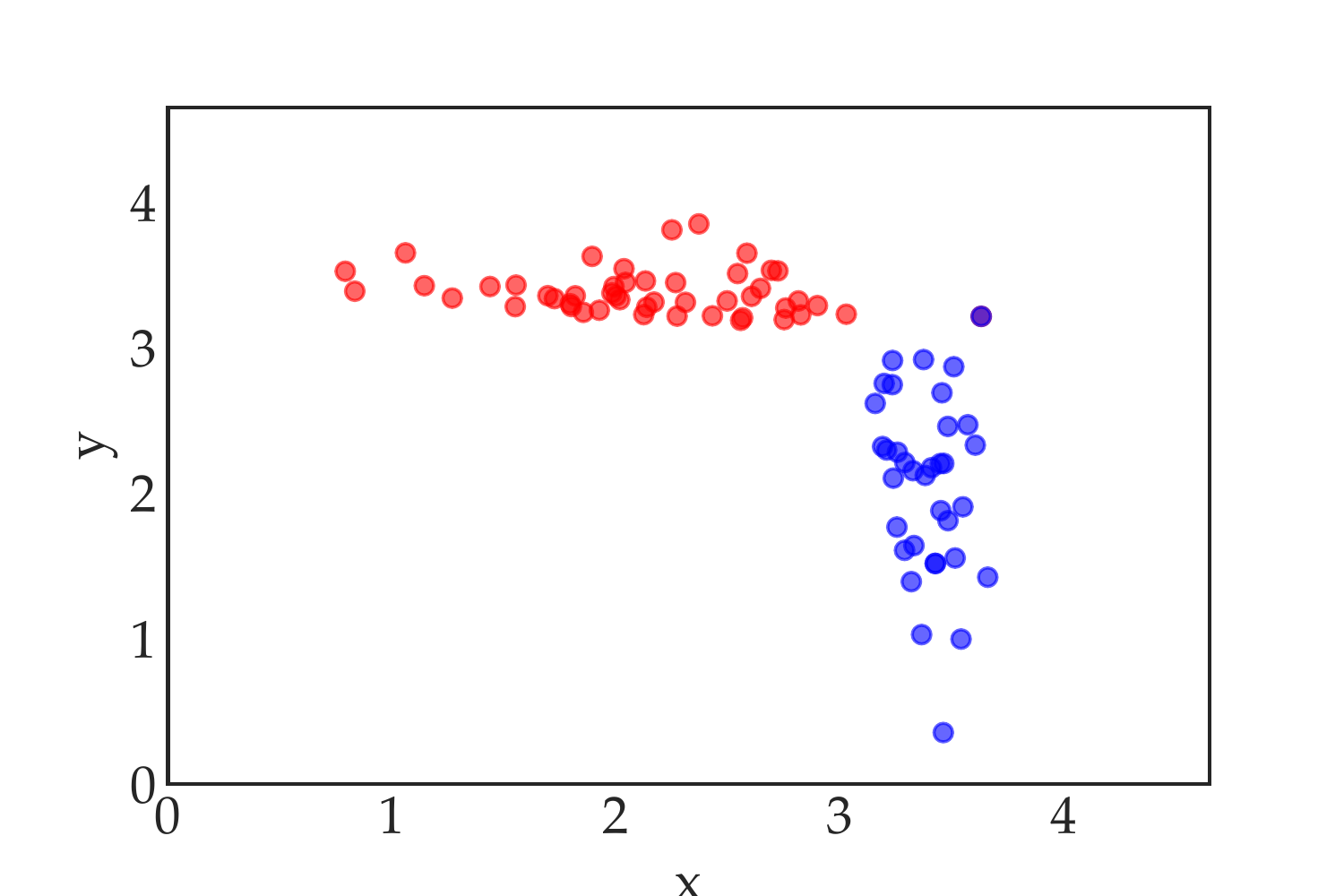}
          \caption{\small{Concentration of excess distribution with $\theta=0.2,0.8$}}
       \end{subfigure}
       \begin{subfigure}{0.42\textwidth}
          \includegraphics[width=0.98\textwidth]{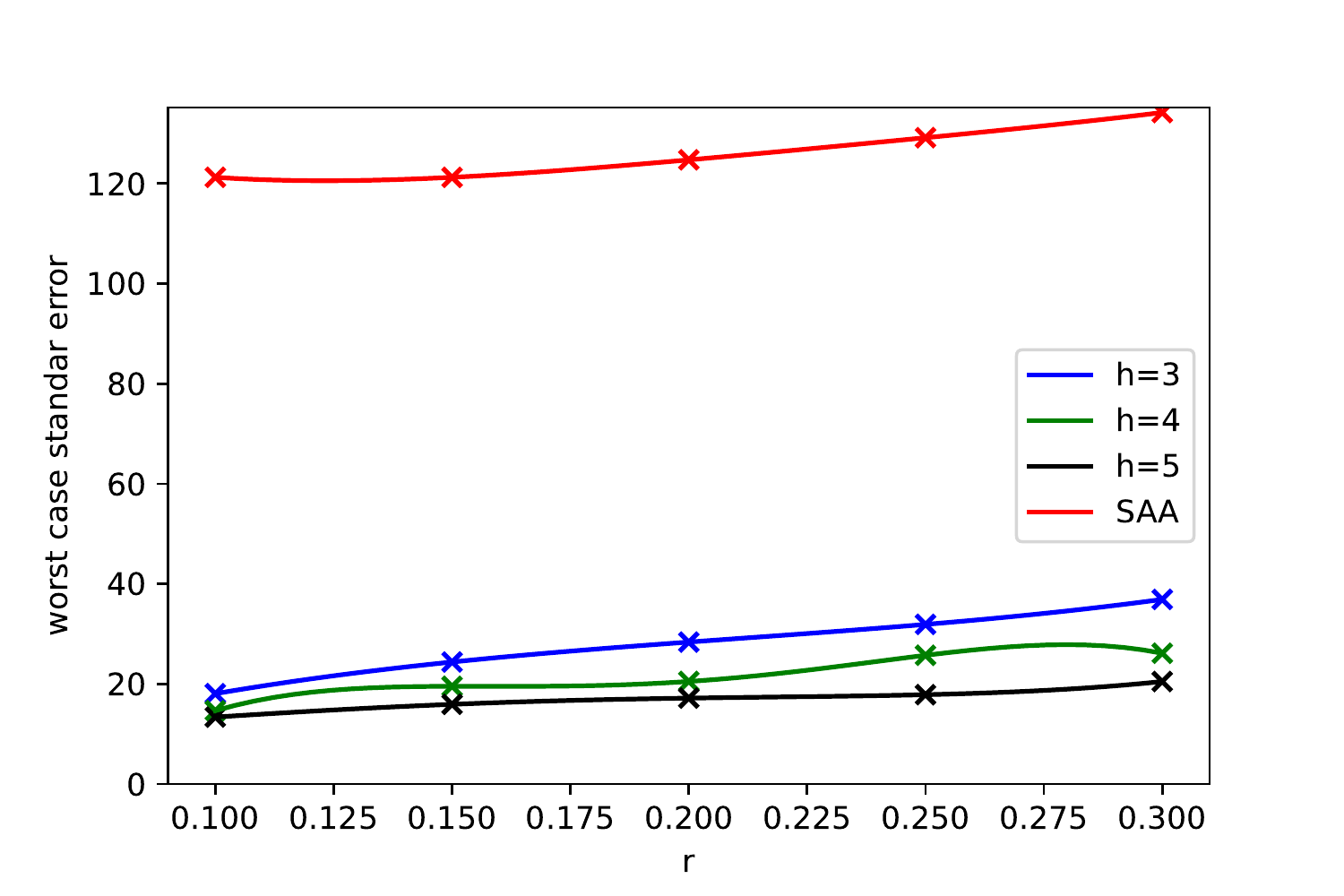}
          \caption{\small{Worst case standard errors over $S_{r}$}}
       \end{subfigure}
      \end{center}
      \caption{{Figures~\ref{fig:Conc-Props}(a) plots the conditional distribution of $\theta X_1 +(1-\theta)X_2$ conditional on it exceeding $3.5$. Figure~\ref{fig:Conc-Props}(b) shows the worst case relative errors (ratio of standard deviation to mean) for the functions $f_{is}(\cdot)$ over $S_{r}$ and compares them with SAA. 
      } \label{fig:Conc-Props} }
 \end{figure}

\begin{remark}\em\label{rem:T-intuition} Solving \eqref{eqn:CVaR_OpT} accurately requires accurate approximation of the objective function $f$ by its IS-weighted sample average $\hat{f}_{is,n}$. To this end, one seeks to find an alternate measure that resembles the conditional distribution of the random vector $\XX$ in region $\{\xx: \ell(\xx,\ttheta) >u\}$. 
 Whenever $\XX$ and the function $\ell(\xx,\ttheta)$ satisfy Assumptions~\ref{assume:Log-Weibull} and \ref{assume:L} respectively, the distribution of $\mv{T}_h(\XX)$ resembles that of $\XX$ in the region $\{\xx: \ell(\xx;\ttheta)>u \}$, irrespective of the choice of decision parameter $\ttheta$ (this can be seen through an application of \citeNP[Proposition 5.1]{deo2021achieving}).  Proposition~\ref{prop:IS-Uniform} showcases that this translates to an exponential reduction in the variance error incurred while estimating the objective function $f$ using the importance weighted sample average $\hat{f}_{is,n}$.
\end{remark}
\begin{example}\label{eg:Example1}\em 
To illustrate the difficulty in applying IS to the optimisation context, and the implications of Proposition~\ref{prop:IS-Uniform}, consider a simple two dimensional setting, where $\ell(\theta,x_1,x_2) = \theta x_1 +(1-\theta)x_2$. Notice that the region of concentration of the conditional distribution changes greatly when $\theta$ is changed from $0.2$ to $0.8$. Therefore, an IS distribution which provides a large variance reduction in the objective for $\theta=0.2$ may not for $\theta=0.8$. However, the uniformity of variance reduction in \eqref{eqn:IS-Uniform} demonstrates that by use of IS transformations this difficulty may be mitigated. For the same linear loss but with $\XX\in \R^5$, Figure~\ref{fig:Conc-Props}(b) shows that the worst case standard error over $S_{r}$ is much smaller for IS than it is for SAA for $\beta=0.001$. This demonstrates that significant benefits are obtained through use of IS transformations for realistic values of $\beta$ (rather than the theorem truly holding only asymptotically).
\end{example}
Proposition~\ref{prop:IS-Uniform} can be used to derive variance guarantees on the error in optimising CVaR with \eqref{eqn:FIS}.

\begin{theorem}\label{thm:CVaR-IS-LogWB}
Under Assumptions \ref{assume:Log-Weibull} and \ref{assume:L}, we have  $\sqrt{n}(\hat{c}_{is,n} -c_\beta)\Rightarrow \sigma_{is}(\beta)N(0,1)$ where
\begin{equation}\label{eqn:lim-variance}
    \sigma_{is}^2(\beta) = \text{var}\left((\ell(\ZZ;\ttheta^*) - v_\beta(\ttheta^*))^+\mathcal{L}_h\right)
\end{equation}
Further, the limiting variance $\sigma^2_{is}(\beta)$ of the IS estimator is smaller than the limiting variance $\sigma_c^2(\beta)$ of the SAA estimator \eqref{eqn:CVaR_OpT} as given by the relationship,   
\[
\frac{\sigma^2_{is}(\beta)}{\sigma^2_{c}(\beta)} =o(\beta^{1-\varepsilon}), \text{ for any }\varepsilon>0,
\]
\end{theorem}
Considering the proposed change of measure for CVaR optimisation, Theorem ~\ref{thm:CVaR-IS-LogWB} guarantees a sample complexity of $o(\beta^{-\varepsilon})$ as $\beta \searrow 0,$ where $\varepsilon > 0$ can be made arbitrarily small. With the variance reduction guarantee holding for any choice of hyper-parameter $h > 0,$ an effective $h$ can be chosen via cross-validation without incurring a change of scaling in sample complexity as demonstrated in Section~\ref{sec:SAA-IS-RA++}.

\section{A VANILLA RETROSPECTIVE APPROXIMATION SCHEME FOR CVAR OPTIMIZATION}\label{sec:RA}
The procedure described in Section~\ref{sec:IS_algo} arrives at the optimal CVaR by generating a \textit{single} sample path $\{\XX_1,\ldots,\XX_n\}$ and then solving the resulting IS-weighted SAA problem. However, in practical implementation, solutions are not available in closed form, and sample path problems such as \eqref{eqn:FIS} need to be solved using deterministic optimisation algorithms. Thus, obtaining a solution to within a small tolerance level may be a computationally challenging task. With this in mind, sequential procedures are often deployed to solve SAA (see \citeNP{bayraksan2011sequential,pasupathy2010choosing}). One such method is Retrospective Approximation (RA), which generates a \textit{sequence} of SAA sub-problems, and progressively increases sample sizes, while solving the resulting SAA sub-problems with progressively reducing error tolerances. Further, the solution of each stage is fed into as the starting point for the optimiser in the subsequent stage. This eases the overall computational burden as follows: the initial stages obtain a rough estimate of the optimal parameters while not expending too much computation. The later stages are accurate, since the initial solution to optimisation algorithm used to solve the SAA sub-problem is close to the true solution.

In our context, given a sequence of sample sizes $\{m_k:k\geq 1\}$, this amounts to minimising a sequence of random functions $\{f_{is, m_k}(u,\ttheta):k\geq 1\}$, where $f_{is,n}(u,\ttheta) \to f(u,\ttheta)$ in probability as $n\to\infty$. Further, at each stage $k$,  we set the initial solution to the $(k+1)th$ sub-problem to be $(u_k,\ttheta_k)$ the minimiser of $f_{m_k}$. Algorithm~\ref{algo:CVaR-I.S-RA} gives a RA based implementation of the IS-weighted SAA algorithm from Section~\ref{sec:IS_algo}. Denote the solution of the sample path problem in \eqref{eqn:CVaR-comp-Is} by $(u_k^*,\ttheta_k^*)$. Notice that this is the \textit{true solution} of the optimisation problem and not the one computed to $\varepsilon_k$ precision. Our first result shows that use of IS significantly reduces sample errors in the solution to the sequence of sub-problems. Define 
\begin{equation}\label{eqn:Grad_Opt}
 \mv{g}(u;\ttheta) = \left[1- \beta^{-1}{E}\left(\mv{I}(\ell(\XX,\ttheta) \geq u)\right), \beta^{-1}E\left(\nabla_{\ttheta} \ell(\XX, \ttheta)\mv{I}(\ell(\XX,\ttheta) \geq u)\right) \right] \text{ and }
\end{equation}
\begin{equation}\label{eqn:Sample_Grad}
    \mathbf{G}(\xx; u,\ttheta,r) = \left[1- \beta^{-1}\mv{I}(\ell(\xx,\ttheta) \geq u) \mathcal{L}_{r}(\xx) \ , \ \beta^{-1}\left(\nabla_{\ttheta} \ell(\xx, \ttheta)\mv{I}(\ell(\xx,\ttheta) \geq u)\right)\mathcal{L}_r(\xx)  \right],
\end{equation}
where, $\nabla_{\ttheta}$ denotes the derivative with respect to $\ttheta$. 
\begin{proposition}\label{prop:CLT-RA}\em
Suppose Assumptions~\ref{assume:Log-Weibull} and \ref{assume:L} are satisfied. Further, let the problem in \eqref{eqn:CVaR_OpT} have a unique minimiser. Then, as $k\to\infty$, we have
\begin{equation}\label{eqn:Retrospective-SAA-CLT}
    \sqrt{m_k}[ (u_{k}^*,\ttheta_{k}^*) - (v_\beta^*(\ttheta),\ttheta^*)] \xrightarrow{L} N(\mv{0}, \Sigma_{h}), \text{ where } \Sigma_{h} =  [\nabla \mv{g}(u^{*};\ttheta^*)]^{-\intercal} \text{var}\left(\mv{G}(\ZZ; u^*,\ttheta^*,h) \right) [\nabla\mv{g}(u^{*};\ttheta^*)]^{-1} 
\end{equation}
Further, the Frobenius norm of the limiting covariance satisfies  $\|\Sigma_{h}\|_F =o(\beta^{-\varepsilon})$ for any $\varepsilon>0$. 
\end{proposition}
Proposition~\ref{prop:CLT-RA} showcases the utility of using IS in conjunction with RA to get a better performance in terms of estimation errors. In particular, the second part above suggests that the sample complexity of estimation grows as $o(\beta^{-\varepsilon})$ rather than $\tilde{O}(\beta^{-1})$.
\begin{algorithm}[h]
 \caption{Retrospective Approximation based CVaR Optimisation -- $\textsc{OPT}(h)$}\label{algo:CVaR-I.S-RA}
  \textbf{Input:} Target tail probability level $\beta$, initial solution $u_0,\ttheta_0$, sample sizes $m_1,\ldots,m_k$, error tolerances $\varepsilon_1,\ldots, \varepsilon_k$.\\
  
 \noindent \textbf{ For $k\geq 1$, do}\\
 
  \textbf{1. Transform the samples:} For each sample
  $i=1,\ldots,m_k,$ compute the transformation,
  \begin{align*}
    \ZZ_{i} = \mv{T}_h(\XX_i) := \XX_i [s_{h}]^{\kappa(\XX_i)},
   \end{align*}
\noindent \textbf{2. Solve the IS based optimisation:}
   \begin{equation}\label{eqn:CVaR-comp-Is}
       \hat{c}_{is,m_k} := \inf_{u,\ttheta}\left[u + \frac{1}{m_k\beta}
      \sum_{i=1}^{m_k} \big(\ell({\mv{Z}}_{i},\ttheta) - u \big)^+
      \mathcal{L}_{h,i},\right] \text{ to a tolerance of } \varepsilon_k
    \end{equation}
      with an initial solution $(u_{k-1},\ttheta_{k-1})$. Return also the optimiser of \eqref{eqn:CVaR-comp-Is}, $(u_k,\ttheta_k)$.
  \end{algorithm}  
Notice that Proposition~\ref{prop:CLT-RA} gives no mention of the computation required to  solve the optimisation problem in Algorithm~\ref{algo:CVaR-I.S-RA}, and instead focuses on the quality of the optimal solutions to the sample path sub-problems. In the next discussion, we argue the computational benefits obtained by the use of retrospective approximation continue to hold even when IS is used. Assumption~\ref{assume:RA-size} below imposes a mild condition  on $m_k,\varepsilon_k$  (see \citeNP[Assumptions C.1-C.3]{pasupathy2010choosing}).
\begin{assumption}
\label{assume:RA-size}\em
Suppose that the sequence $\{(\varepsilon_k,m_k): k \geq 1\}$ satisfy the following requirements: 
\begin{enumerate}
    \item If the optimisation procedure used to solve the individual sample path problems exhibits linear convergence, $\liminf_{k\to\infty}\varepsilon_{k-1} \sqrt{m} >0$. If this procedure exhibits polynomial convergence, $\liminf_{k\to\infty} {\log 1/\sqrt{m_{k-1}}}({\log \varepsilon_k})^{-1} >0$.
    \item $\limsup_{k\to\infty} (\sum_{j=1}^k m_j)^2/\varepsilon_{k}^2<\infty $
    \item $\limsup_{k\to\infty} m_k^{-1}\sum_{j=1}^k m_j <\infty$.
\end{enumerate}
\end{assumption}
Observe that there are two competing errors in the solution output by Algorithm~\ref{algo:CVaR-I.S-RA}. First, there is the sample error, which can be quantified by the limit theorem in Proposition~\ref{prop:CLT-RA}. Second, there is the error due to solving the optimisation problem imperfectly (up to $\varepsilon_k$ accuracy). Assumption~\ref{assume:RA-size} imposes conditions so that these errors are balanced out and an optimal rate of convergence of the solution is obtained.  Notice that the total work done in running Algorithm~\ref{algo:CVaR-I.S-RA} for $k$ epochs equals $W_k = \sum_{j\leq k} N_j m_j$, where $N_j$ is the (random) number of calls made to the deterministic algorithm used to solve \eqref{eqn:CVaR-comp-Is} to $\varepsilon_j$ accuracy. Proposition~\ref{prop:IS-RA-work-error}  establishes that the work normalised error for the RA based IS-weighted CVaR optimisation remains bounded.
\begin{proposition}\label{prop:IS-RA-work-error}\em 
Under Assumptions~\ref{assume:Log-Weibull} - \ref{assume:RA-size}, we have $W_k [\sigma_{is}(\beta)]^{-2}\|(u_k,\ttheta_k) - (v_\beta(\ttheta^*),\ttheta^*)\|_2^2 =O_p(1)$.
\end{proposition}
Notice the appearance of the limiting variance $\sigma_{is}(\beta)$ from \eqref{prop:IS-RA-work-error} above. This suggests that the IS scheme continues to enjoy a significant reduction over SAA in work normalised errors. Indeed this is showcased in experiments in Section~\ref{sec:numbers}, where we show that the amount of computing effort required to solve the CVaR optimisation problem to a fixed accuracy is substantially less when IS is used.  

\section{ENHANCED RETROSPECTIVE APPROXIMATION WITH ADAPTIVITY IN IS CHOICE}\label{sec:SAA-IS-RA++}
One major shortcoming of Algorithm~\ref{algo:CVaR-I.S-RA} is the lack of optimisation over the cross-validation parameter $h$. In this discussion, we develop an enhanced version of the retrospective approximation based CVaR optimisation which includes a subroutine to iteratively arrive at a value of $h$ which gives a lower error. To simplify matters, we assume the existence of a noisy oracle, which given $n$ samples of data and $(u,\ttheta)$ returns a value $\hat{h}_n(u,\ttheta)$, such that $\hat{h}_n(u_n,\ttheta_n) \xrightarrow{P} h(u,\ttheta)$ whenever $(u_n,\ttheta_n) \to (u,\ttheta)$ as $n\to\infty$. We further assume that $h(u^*,\ttheta^*)$ is a good hyper-parameter choice for problem (the specific implementation for $h_n$ and $h$ is presented in Section~\ref{sec:numbers}). Under this assumption, Algorithm~\ref{algo:CVaR-FULL.} presents an enhancement to the RA based IS-weighted CVaR optimisation scheme.
\begin{algorithm}[h]
 \caption{Enhanced Retrospective Approximation based CVaR Optimisation}\label{algo:CVaR-FULL.}
  \textbf{Input:} Target tail probability level $\beta$,  samples 
  $\boldsymbol{X}_1,\ldots,$ from $f_{\XX}(\cdot)$, initial seeds $u_0,\ttheta_0,h_0$.\\
 
 \noindent\textbf{ For $k\geq 1$, do}\\
 
  \textbf{1. IS-Weighted CVaR optimisation:} With a sample size of $m_{k,1}$ and error tolerance $\varepsilon_{k,1}$, implement steps 1 and 2 of Algorithm~\ref{algo:CVaR-I.S-RA} starting from $(u_{k-1},\ttheta_{k-1})$ and with $h=h_{k-1}$.\\
\noindent \textbf{2 Update the cross validation parameter: }  Using a sample size of $m_{k,2}$, return the updated value $h_{k} = \hat{h}_{m_{k,2}}(u_k,\ttheta_k)$
  \end{algorithm}  
It possesses the following two benefits over prominent adaptive IS schemes:

For every value of $\{h(u,\ttheta):u >0,\ttheta\in \Theta\}$, the distribution for $\ZZ$ may be generated easily by suitably transforming the samples generated from the distribution of $\XX$. The ability to cheaply obtain samples from an alternate distribution, as assumed with existing IS techniques, could be an issue when the IS distribution family need to be expressive enough to approximate the zero variance measures for complex optimization objectives. 

Similarly, given a value of $(u,\ttheta)$, Algorithm~\ref{algo:CVaR-FULL.} deals with the selection of $h(u,\ttheta)$ directly from samples of $\XX$, rather than assuming a black box access to a good choice of $h(u,\ttheta).$ For a fixed choice of $u,\ttheta,$ the selection of $h(u,\ttheta)$ is fortunately a one-dimensional problem which can be tackled effectively with bisection or grid search and using common random  numbers. Using the same samples of $\XX$ to reduce the variance in IS parameter selection is not possible with most existing IS approaches. As will be explained shortly, the computed values of the hyper-parameter will converge in probability to $h(u^*,\ttheta^*)$. Since the entire cross validation procedure only requires access to samples of $\XX$, it is more easily implementable than adaptive techniques which require black box access to a good importance sampler at each stage.

Observe that  given the $(k-1)$th stage solution, $(h_{k-1},u_{k-1},\ttheta_{k-1})$ and $m_{k,1}$ number of samples, as a consequence of the discussion in \shortciteNP[Theorem 2]{pasupathy2010choosing}, $(u_k,\ttheta_k) = (u^*,\ttheta^*) + O_{p}(m_{k,1}^{-1/2}+\varepsilon_k)$.  Further, notice that with $m_{k,2}$ samples used for cross-validation in the first stage, one has $\hat{h}_{m_{k,2}}(u_{k-1},\ttheta_{k-1}) = h(u_k,\ttheta_k)+o_{{p}}(1)$. From the continuity of $h(\cdot)$ in the parameter, and the consistency of $(u_{k-1},\ttheta_{k-1})$ as $k\to\infty$, we further have that $\hat{h}_{m_{k,2}}(u_{k-1},\ttheta_{k-1}) = h(u^*,\ttheta^*)+o_{{p}}(1)$. The above discussion suggests that Algorithm~\ref{algo:CVaR-FULL.} arrives at a statistically consistent estimate of the desired optimal hyper-parameter. It is of note to observe that the convergence to the optimal $h(u^*,\ttheta^*)$ appears to be insensitive to the choice of $h_0$. Indeed, this is  demonstrated experimentally in Section~\ref{sec:numbers}.

A natural question to ask given the above discussion, is whether one may derive a limit theorem which states that as $k\to\infty$, $\sqrt{m}_k(\hat{c}_{is,m_k} -c_\beta) \xrightarrow{L} N(0,[\sigma_{is}^*(\beta)]^2)$, where $[\sigma_{is}^*(\beta)]^2$ is the variance of the objective evaluated at $(u^*,\ttheta^*)$ under the distribution $\ZZ_{h(u^*,\ttheta^*)}$. Addressing this is an interesting follow-up direction. 
We next attempt to justify the benefits obtained by the use of Algorithm~\ref{algo:CVaR-FULL.} through numerical experiments.

\section{NUMERICAL EXPERIMENTS}\label{sec:numbers}
\subsection{Minimization of CVaR for Linear Portfolios} 
\begin{figure}[t]
      \begin{center}
      \begin{subfigure}{0.45\textwidth}
          \includegraphics[width=0.98\textwidth]{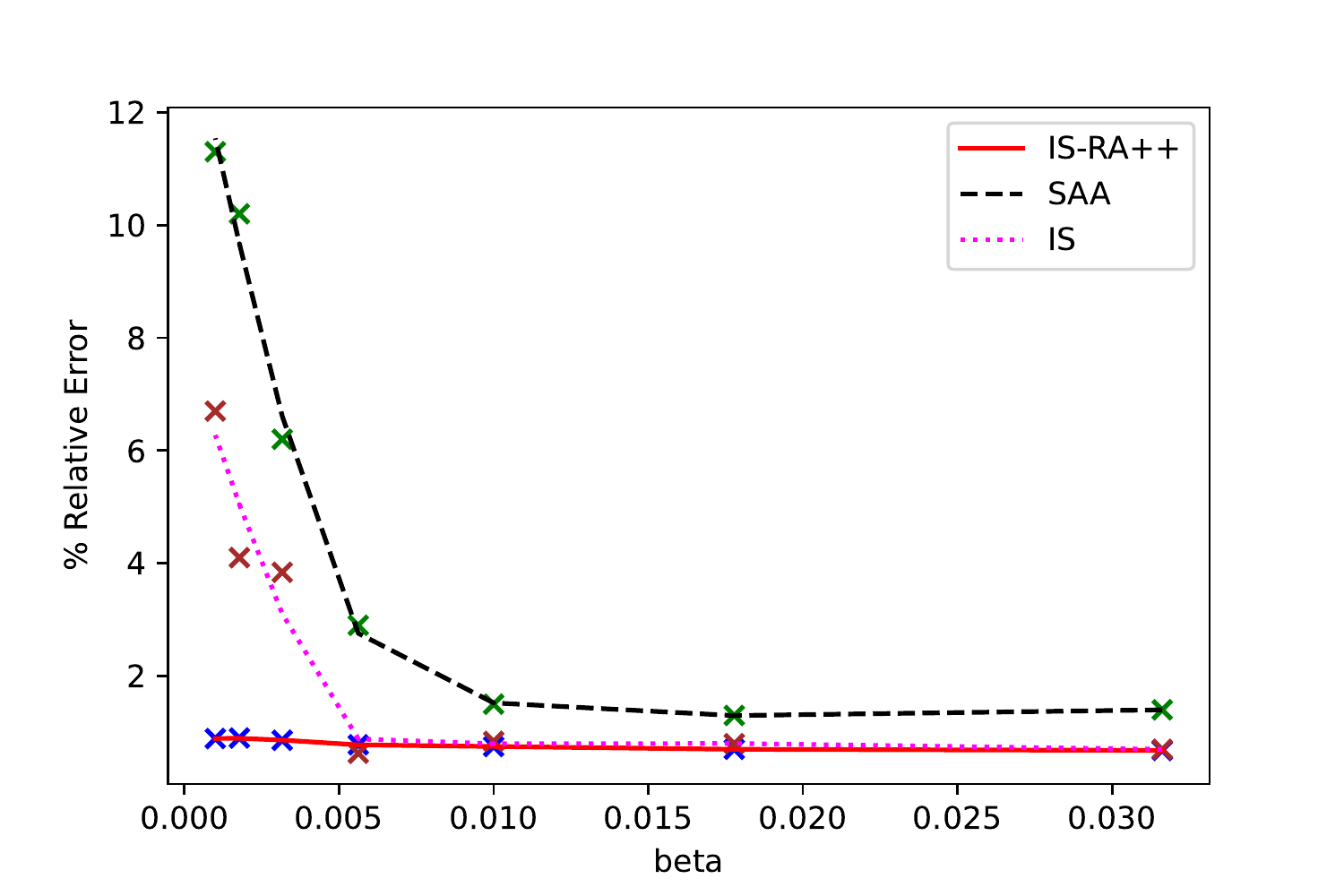}
          \caption{\small{Relative RMSE  in CVaR optimisation}}
       \end{subfigure}
       \begin{subfigure}{0.45\textwidth}
          \includegraphics[width=0.98\textwidth]{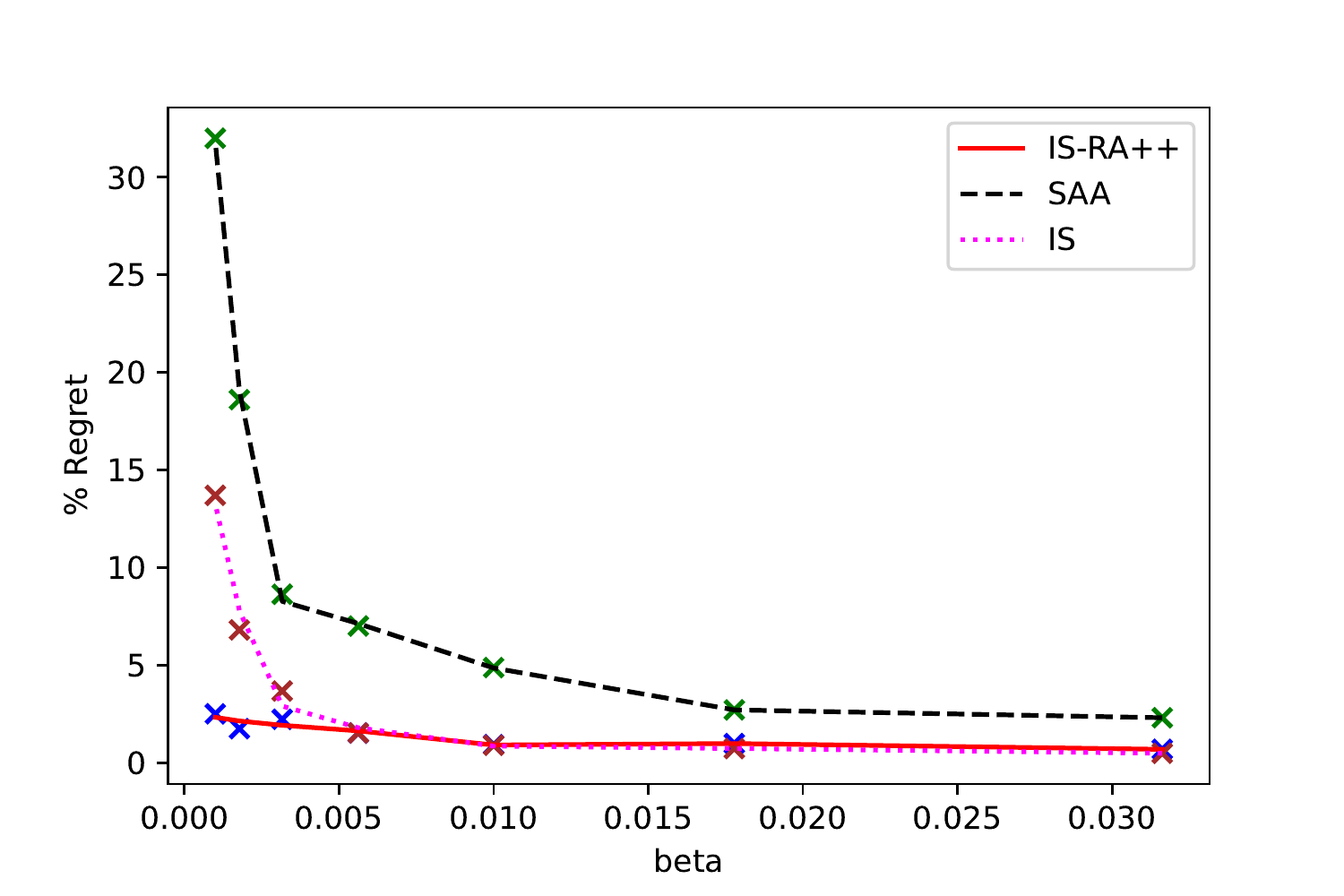}
          \caption{\small{Relative regret in CVaR optimisation}}
       \end{subfigure}
      \end{center}
      \caption{{Figure~\ref{fig:RMSE-NN}(a) compares the relative RMSE in CVaR optimisation using Algorithms~\ref{algo:CVaR-I.S-RA} and \ref{algo:CVaR-FULL.} with plain SAA. Figure~\ref{fig:RMSE-NN}(b) compares the corresponding out-of-sample regret. In each of the figures, cross-marks respectively denote estimated quantities. 
      } \label{fig:RMSE-NN} }
 \end{figure}

In order to demonstrate the efficacy of our algorithms in practice, we consider solving the constrained minimum CVaR minimisation portfolio optimisation problem. In this setting, $\ell(\xx,\ttheta) = \ttheta^\intercal\xx $, and the set $\Theta = \{\ttheta:\ttheta^{\intercal}\mv{1} = 1\}$.
The marginals of $\XX$ are taken to have the c.d.f.s $F_i(x) = P(X_i \leq x) = 1-\e^{-x^{\alpha_i}}$ where $\alpha_i =0.5 \ \forall i$.  Dependence is modelled through a Gaussian copula whose covariance matrix $\mv{R}$ is designed to capture a realistic degree of correlation among various asset returns. We compare the effectiveness of the estimators returned by Algorithms~\ref{algo:CVaR-I.S-RA} and \ref{algo:CVaR-FULL.} with plain SAA using three performance measures:

First, we use relative mean square error as a metric to compare the quality of the optimal value output by each of the algorithms. That is, given the output of an algorithm, call it $\hat{c}_{\beta}$, we compute the relative root mean square error as 
$\sqrt{\frac{1}{K}\sum_{i=1}^K( {\hat{c}_{\beta,i}}/{c_\beta} -1)^2}${ where } $\hat{c}_{\beta,i}$ { are outputs of independent runs of the algorithm and $c_\beta$ is defined in \eqref{eqn:CVaR_OpT}}.
We use relative errors as metrics of performance both here as well in the next experiment to facilitate a scale free (i.e: $\beta$ independent) comparison between different algorithms. In order to compute the optimal value $c_\beta$, we solve the SAA problem \eqref{eqn:PMC} with $N=10^6$ samples. In each case, the value of $\beta$ is varied from $\beta=0.037$ to $\beta=5\times 10^{-4}$. In order to ensure a fair comparison between each of the three methods, we keep the total sample budget fixed and equal to $2500$. For the implementation in Algorithm~\ref{algo:CVaR-FULL.}, this budget is divided among two epochs as $m_1=500$ and $m_2=2000$. Further, for every $(u,\ttheta)$, we let $\hat{h}_{n}(u,\ttheta) = \inf_{h}\frac{1}{n} \sum_{i=1}^n \mathbf{I}(\ell(\ZZ_{h,i};\ttheta ) \geq u) \mathcal{L}_{h,i}^2$.
Observe that from the law of large numbers this converges to the second moment of the gradient of the objective function with respect to $u$. Since the final errors in decision (see Proposition~\ref{prop:CLT-RA} for example) depend on this quantity having a smaller variance, $\hat{h}_n(u,\ttheta)$ as chosen above is reasonable. For Algorithm~\ref{algo:CVaR-I.S-RA} for all $\beta$, we choose $h=2.5$.  Figure~\ref{fig:RMSE-NN}(a) details the results. 

\begin{figure}[t]
      \begin{center}
      \begin{subfigure}{0.42\textwidth}
          \includegraphics[width=0.98\textwidth]{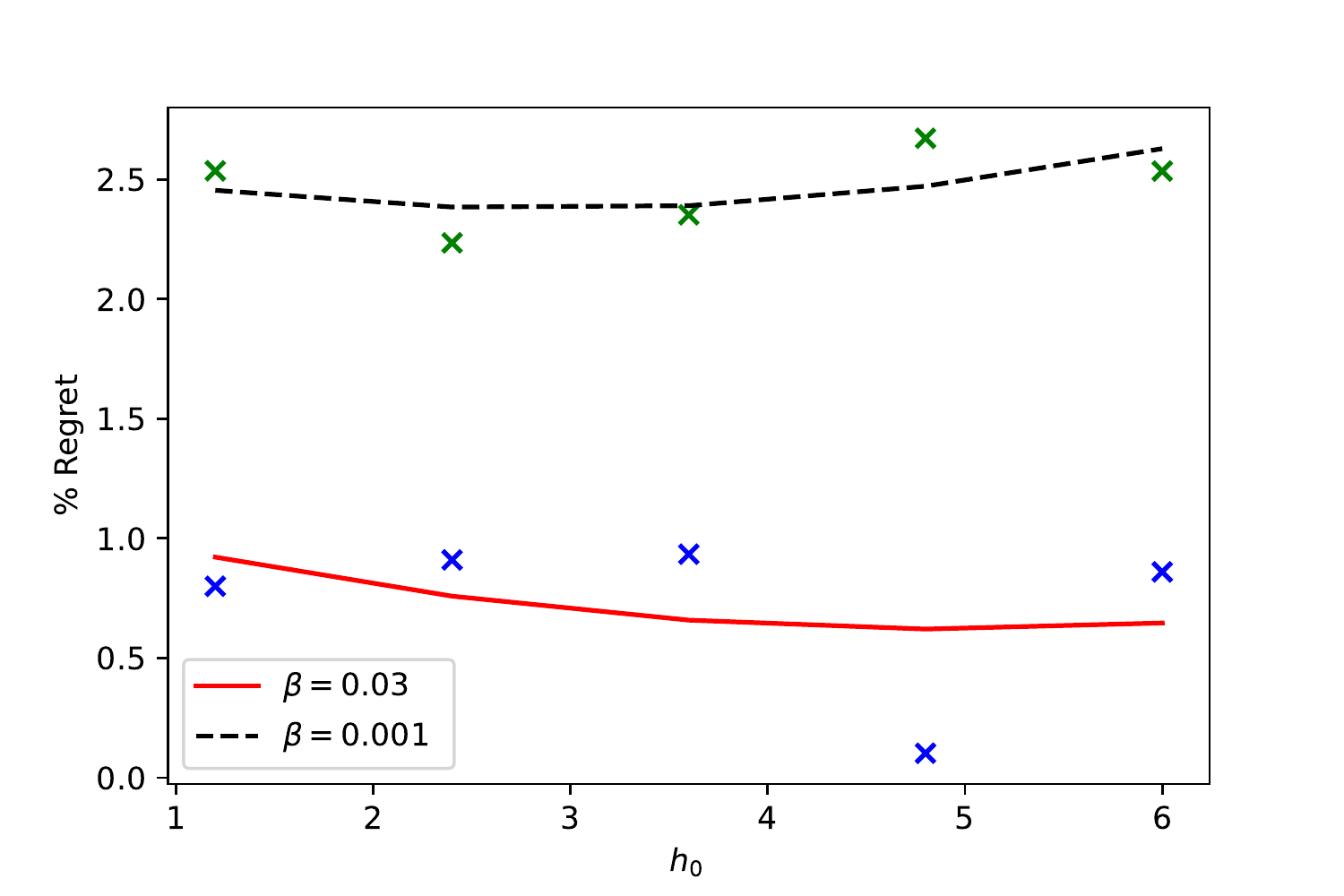}
          \caption{\small{Relative regret as a function of $h_0$}}
       \end{subfigure}
       \begin{subfigure}{0.42\textwidth}
          \includegraphics[width=0.98\textwidth]{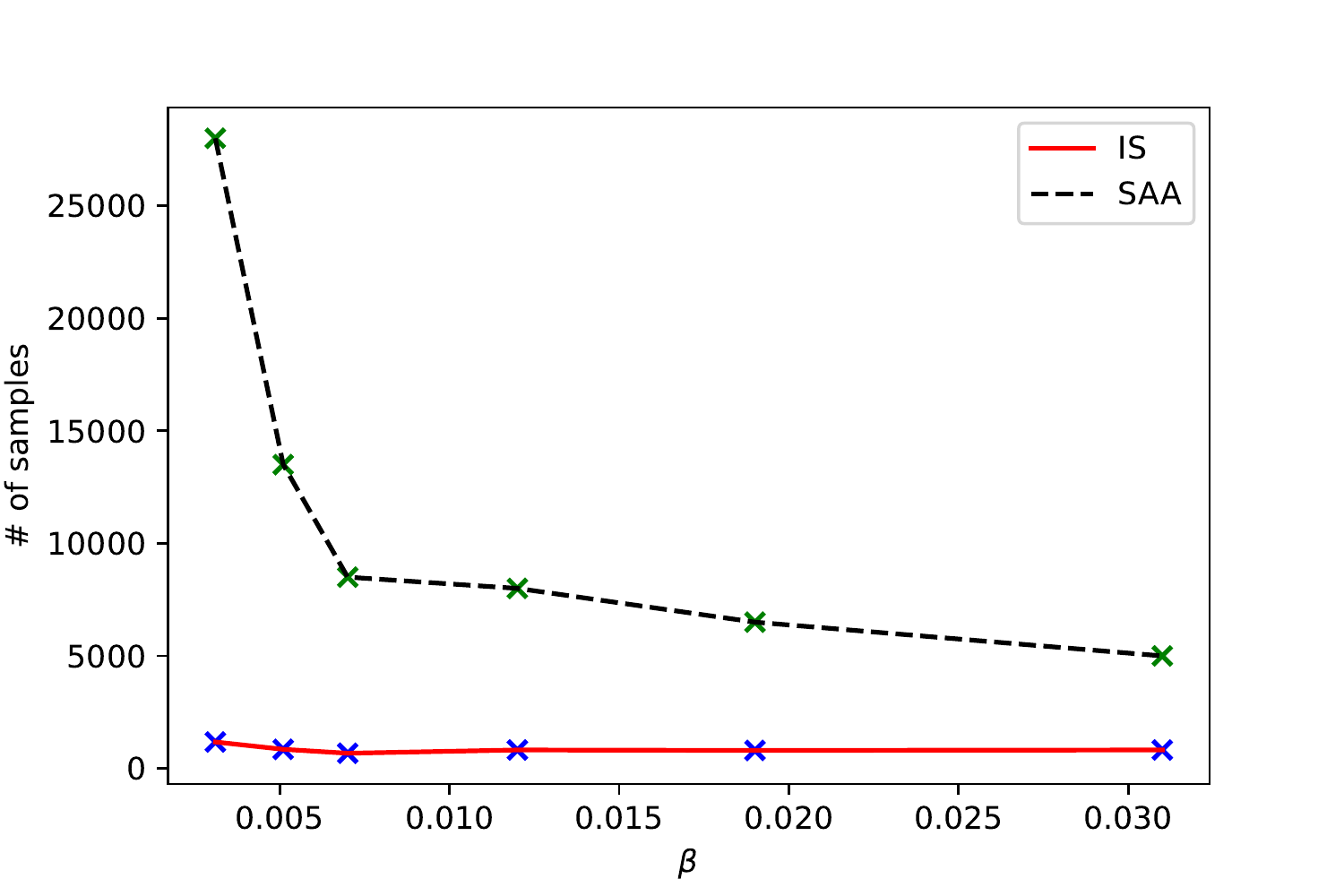}
          \caption{\small{Samples required to obtain $1\%$ relative regret}}
       \end{subfigure}
      \end{center}
      \caption{{Figure~\ref{fig:Regret-Robustness}(a) plots the relative regret incurred in optimisation as a function of the initial value $h_0$. Figure~\ref{fig:Regret-Robustness}(b) plots the number of samples required by each method to obtain $1\%$ relative regret. In each of the figures, cross-marks respectively denote estimated quantities. 
      }  }\label{fig:Regret-Robustness}
 \end{figure}

We next use relative regret as a metric to compare out-of-sample performance each of the algorithms. Given the output of an algorithm $\hat{\ttheta}$, we define the relative regret as 
$\hat{r}_\beta(\hat{\ttheta}) = \frac{1}{K}\sum_{i=1}^K({c_\beta(\hat{\ttheta}_i)}/{c_\beta} - 1) \times 100\%$, { where } $\hat{\ttheta}_{i}$ { are outputs of independent runs of the algorithm.}  The first observation is that IS-RA significantly outperforms SAA in terms of relative errors obtained. Secondly, adaptively optimising over $h$ as in Algorithm~\ref{algo:CVaR-FULL.} leads to a further improvement over Algorithm~\ref{algo:CVaR-I.S-RA}.
Finally, figure~\ref{fig:Regret-Robustness}(a) displays the relative regret incurred as a function of the starting point of the $h_0$. Notice that there appears to be a degree of robustness in the performance of the algorithm to the specific value of $h_0$ selected.

Finally, in order to compare in terms of the effort required to obtain a desired out of sample accuracy, we compute the number of samples required by each method to obtain $1\%$ relative regret; refer Figure\ref{fig:Regret-Robustness}(b). Observe that for $\beta=0.037$, for IS, this is $\approx 600$, while SAA requires $\approx 5500$ samples. This difference is even more pronounced when $\beta= 0.003$, where SAA requires roughly $28000$ samples, while IS only requires $1175$.   
\subsection{A Portfolio Credit Risk Example} 
\begin{figure}[t]
      \begin{center}
          \includegraphics[width=0.42\textwidth]{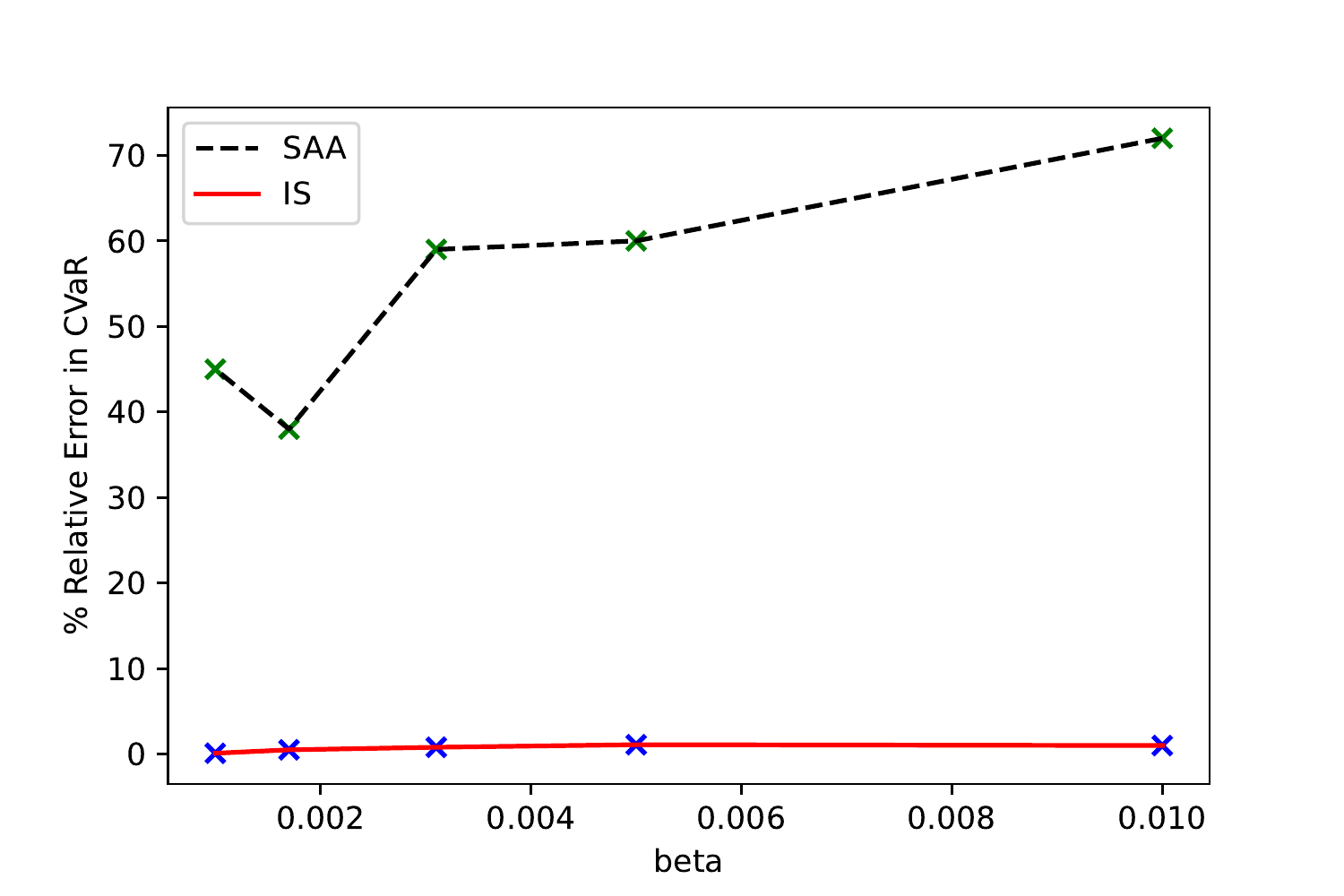}
      \end{center}
\caption{\small{Error in CVaR Computation for the portfolio credit risk problem}}\label{fig:PCR-VaR}
 \end{figure}

Consider the problem of selecting a portfolio of loans from among $K$ classes (industries). Given a that a set of common market factors $\XX$ realise a value $\xx$, a loan belonging to class $i$ defaults with probability $p_i(\xx)$ independently of everything else. We further suppose that the loss given default for loan $k$ is given by $e_k$ (random), independent of the market variable. Denoting $n_i$ as the number of loans belonging to class $i$, the total loss given a portfolio $\ttheta$ is $\ell(\ttheta,\XX) = \sum_{i=1}^K \theta_i ( \sum_{k=1}^{n_k} e_k \mathbf{I}(\text{loan k defaults}))$. The objective is to find the portfolio with the minimum CVaR, subject to the total returns from the investment exceeding a nominal threshold, that is $\ttheta^\intercal\mv{r} \geq q$ where the portfolio returns are given by $\mv{r}$. In this example, $K=2$ and $\XX\in \R^4$ has Weibull marginals with a Gaussian copula used to model joint dependence. We further assume that $p_{i}(\xx)$ has a logistic form. Importance sampling is deployed only on the common market factors $\XX$, and the IS weighted algorithm is solved as before. In this implementation, $N=2000$ while $n_1=n_2=5000$ and $e_i$ are uniform random variables. Figure~\ref{fig:PCR-VaR} displays the results of the experiment. Notice that the use of IS on the common market variable significantly improves lowers errors in optimal VaR estimation. The relative error in CVaR estimation is roughly $1\%$ when IS is used as opposed to over $50\%$ without the use of IS.
\section{OUTLINES OF KEY PROOF STEPS}\label{sec:proofs}

\noindent \textbf{Proof of Proposition~\ref{prop:IS-Uniform}:} 
For notational convenience, let $M_{2,u}$ denote the second moment of $[\mathcal{L}_h(\ell(\ZZ,\ttheta) - u)^+]$. Note that this is an upper bound for the variance which we want to compute. Further, define $t(u) = \Lambda_{\min}(u^{1/\rho})$, $q_{\infty}(u)$ as the inverse of $\Lambda_{\min}(u)$
and $\YY_u=  [t(u)]^{-1}\mv{\Lambda}(\XX)$. Changing variables from $\XX $ to $\YY_u $ in the expectation below  (see (EC.16) onward in the proof of Lemma EC.6 of \citeNP{deo2021achieving} for detailed steps in a similar change of variables exercise), we obtain  $M_{2,u}= \Expc \left[(\ell(\XX,\ttheta) - u)^2 \frac{f_{\XX}(\XX)}{{f}_{\XX}(\mv{T}^{-1}_h(\XX))} J(\mv{T}^{-1}_h(\XX)); \ell(\XX,\ttheta) \geq u \right]$. This simplifies to  $\Expc\left[\exp(-t(u) F_u(\YY_u))\right]$ where $F_u(\pp) = a_{u}(\pp) + b_{u}(\pp) +c_u(\pp) - 2d[t(u)]^{-1}\log t(u) + \chi_{\Lev_1^+(\ell_{u,\ttheta})}(\pp)$ where
    $a_u(\pp) = [t(u)]^{-1} \left[ \log f_{\YY}(\mv{\psi}_u(t(u)\pp)    - \log f_{\YY}(t(u)\pp) \right]$, 
 $b_u(\pp) = [t(u)]^{-1} \left[\sum_{i=1}^d \left[
    \log\lambda_i(\mv{T}^{-1}_{h,i}(\mv{q}(t(u)\pp))) -
    \log\lambda_i(q_i(t(u)p_i))\right] -
    \log J_h(\mv{T}^{-1}(\mv{q}(t(u)\pp)))\right]$  and  $ c_u(\pp) = -2[t(u)]^{-1} \log(\ell(\qq(t(u) \pp,\ttheta ) -u))$.
Define $p_u=P(\ell(\XX,\ttheta) \geq u)$. Then an application of the general Varadhan's integral lemma \cite[Theorem 2.1]{LD_Varadhan}, for all $(u,\ttheta)$, we have that $M_{2,u} = o(p_u^{2-o(1)})$ (see the proof of \citeNP[Theorem 1]{deo2021efficient}). Similarly, we have that $\text{var}[(\ell(\XX,\ttheta)-u) \mathbf{I}(\ell(\XX,\ttheta) \geq u)] = \Theta(p_{u})$. Combining these two proves the claim for a fixed $(u,\ttheta)$. Notice that $\inf\{u: u\in S_r\} \geq\log^{\varepsilon}1/\beta$ for all $\beta$ small enough. Further, $\inf_{h>h_{\min}}s_{h} = o(-\log^{\varepsilon} \beta)$. The uniformity over $h,u,\ttheta$ is now obtained using the uniform convergence of $\ell$ as in Assumption~\ref{assume:L} and then following the proof of \citeNP[Lemma D.2]{deo2021achieving}.  For part 2, notice that from the above discussion, for any $(u,\ttheta)$, the ratio of the variances of $\hat{f}_{is,1}$  and  $\hat{f}_1$  is given by $o(p_u^{1-\varepsilon})$. Under Assumptions~\ref{assume:Log-Weibull} and \ref{assume:L}, it can be established that for some continuous function $a(\cdot)$, $C_{\beta}(\ttheta) = a(\ttheta)(1+o(1))q_{\infty}^{\rho}(-\log\beta)$. As a consequence of \citeNP[Theorem 7.33]{rockafellar2009variational}, $\ttheta^*$ which minimises $C_\beta(\ttheta)$ satisfies $\ttheta^* = \ttheta_1+o(1)$, where $\ttheta_1\in\arg\min a(\ttheta)$. Then, the optimal $u^*=v_{\beta}(\ttheta^*) = I^* q_{\infty}^{\rho}(-\log \beta)[1+o(1)]$, where $I^* = \inf_{\ttheta}a(\ttheta)$. This implies that  $u\in S_r$, $u\geq (1-r) I^*q_{\infty}^{\rho}(-\log \beta )$. Finally, it can be seen that $\log p_u = \Lambda_{\min}(u/a(\ttheta))(1+o(1))$ (apply \citeNP[Theorem 4.2]{deo2021achieving}). Plugging everything together, we have that $\log p_u \leq (1-r)^{\alpha_{\min}}( I^*/a(\ttheta))\log \beta $. Now, the continuity of $a(\cdot)$ implies that whenever $(u,\ttheta)\in S_r$, and $h>h_{\min}$, $p_u \leq \beta^{1/1-\rho(r)}$, for $\rho(r) \to 0$. \qed

\noindent \textbf{Proof of Theorem~\ref{thm:CVaR-IS-LogWB}: } Notice first that the limiting function in the optimisation, $f$ may be written as the expected value of $\hat{f}_{is,1}$. Further, observe that due to Assumptions \ref{assume:Log-Weibull} and \ref{assume:L} , $\hat{f}_{is,1}$ has a finite variance. Therefore, by an application \cite[Theorem 3.2]{shapiro1991asymptotic}, one obtains the first part of Theorem~\ref{thm:CVaR-IS-LogWB}. The quantification of the variance reduction is obtained following the proof of Proposition~\ref{prop:IS-Uniform}, with $p_{v_{\beta(\ttheta^*)}} = \beta$.\qed

\noindent \textbf{Proof of Proposition~\ref{prop:CLT-RA}} It is sufficient to verify the conditions from \citeNP[Theorem 4]{pasupathy2010choosing}. First, note that the optimal solutions  $(u_k^*,\ttheta_k^*)$ are zeros $\nabla \hat{f}_{is,m_k}(u,\ttheta)$. Further, we have that $E\mv{G}(u,\ttheta) = g(u,\ttheta)$ for all $(u,\ttheta)$. Finally, owing to Assumptions~\ref{assume:Log-Weibull} and \ref{assume:L}, the usual conditions required for a CLT hold, and we therefore have $\sqrt{m_k}(\nabla \hat{f}_{is,m_k}(u,\ttheta) - \mv{g}(u,\ttheta) ) \to N(0,\text{var}(\mv{G}(\ZZ; u,\ttheta;h)))$. By the assumptions of the theorem, there exists a unique minimiser, and the gradients of $\mv{g}$ is non-singular. Finally, the uniform convergence assumption holds as a result of the discussion following the statement of \citeNP[Theorem 4]{pasupathy2010choosing}. \qed

\noindent \textbf{Proof of Proposition~\ref{prop:IS-RA-work-error}} This follows from Assumption~\ref{assume:RA-size}, Proposition~\ref{prop:CLT-RA} and the proof of \citeNP[Theorem 5]{pasupathy2010choosing}.\qed
\appendix
\section{Expression for the Jacobian $J_h$}  Recall that  $ J_h(\cdot) $ is the Jacobian of the transformation
  $\mv{T}_h(\cdot)$. This is given by
   \begin{equation}
    \label{eqn:Jac}
    J_h(\xx)  := \left[\prod_{i=1}^d \tilde{J}_i(\xx) \right]\times \frac{s_h^{\mv{1}^\intercal \mv{\kappa}(\xx)}}{\max_{i=1,\ldots,d} \tilde{J}_i(\xx)} \text{ where } \tilde{J}_i(\xx)
            := 1+\frac{\rho^{-1}\log(s_h)}{\Vert\log(1+|\xx|)
              \Vert_\infty}\frac{|x_i|}{1+|x_i|}, \quad i = 1,\ldots,d.
              \nonumber
  \end{equation}


\section*{ACKNOWLEDGEMENTS} 
The authors acknowledge support from Singapore Ministry of Education grant MOE2019-T2-2-163. 

{ \small \bibliographystyle{wsc}
\bibliography{demobib}}

 \section*{AUTHOR BIOGRAPHIES}
{\footnotesize  {\bf ANAND DEO} is a Postdoctoral Researcher at Singapore University of Technology and Design. His research interests are in quantitative risk management, operations research and machine learning. His e-mail address is  \url{deo\_avinash@sutd.edu.sg}.}

\noindent {\footnotesize  {\bf KARTHYEK MURTHY} is an Assistant Professor in Singapore University of Technology and Design. His research centers around building models and methods for incorporating competing considerations such as risk, robustness, and fairness in data-driven optimization problems affected by uncertainty.  
His e-mail address is {\url{karthyek\_murthy@sutd.edu.sg}}.}

\noindent {\footnotesize  {\bf TIRTHO SARKER} is a Research Assistant at Singapore University of Technology and Design. His e-mail address is \url{tirtho_sarker@sutd.edu.sg}.}

\end{document}

%% file: wscbib.tex
\makeatletter
\let\@internalcite\cite
\def\cite{\def\@citeseppen{-1000}%
    \def\@cite##1##2{(##1\if@tempswa , ##2\fi)}%
    \def\citeauthoryear##1##2##3{##1 ##3}\@internalcite}
\def\citeNP{\def\@citeseppen{-1000}%
    \def\@cite##1##2{##1\if@tempswa , ##2\fi}%
    \def\citeauthoryear##1##2##3{##1 ##3}\@internalcite}
\def\citeN{\def\@citeseppen{-1000}%
    \def\@cite##1##2{##1\if@tempswa, ##2)\else{}\fi}%
    \def\citeauthoryear##1##2##3{##1 (##3)}\@citedata}
\def\citeA{\def\@citeseppen{-1000}%
    \def\@cite##1##2{(##1\if@tempswa , ##2\fi)}%
    \def\citeauthoryear##1##2##3{##1}\@internalcite}
\def\citeANP{\def\@citeseppen{-1000}%
    \def\@cite##1##2{##1\if@tempswa , ##2\fi}%
    \def\citeauthoryear##1##2##3{##1}\@internalcite}
\def\shortcite{\def\@citeseppen{-1000}%
    \def\@cite##1##2{(##1\if@tempswa , ##2\fi)}%
    \def\citeauthoryear##1##2##3{##2 ##3}\@internalcite}
\def\shortciteNP{\def\@citeseppen{-1000}%
    \def\@cite##1##2{##1\if@tempswa , ##2\fi}%
    \def\citeauthoryear##1##2##3{##2 ##3}\@internalcite}
\def\shortciteN{\def\@citeseppen{-1000}%
    \def\@cite##1##2{##1\if@tempswa, ##2\else{}\fi}%
    \def\citeauthoryear##1##2##3{##2 (##3)}\@citedata}
\def\shortciteA{\def\@citeseppen{-1000}%
    \def\@cite##1##2{(##1\if@tempswa , ##2\fi)}%
    \def\citeauthoryear##1##2##3{##2}\@internalcite}
\def\shortciteANP{\def\@citeseppen{-1000}%
    \def\@cite##1##2{##1\if@tempswa , ##2\fi}%
    \def\citeauthoryear##1##2##3{##2}\@internalcite}
\def\citeyear{\def\@citeseppen{-1000}%
    \def\@cite##1##2{(##1\if@tempswa , ##2\fi)}%
    \def\citeauthoryear##1##2##3{##3}\@citedata}
\def\citeyearNP{\def\@citeseppen{-1000}%
    \def\@cite##1##2{##1\if@tempswa , ##2\fi}%
    \def\citeauthoryear##1##2##3{##3}\@citedata}
%
%
%
\def\@citedata{%
    \@ifnextchar [{\@tempswatrue\@citedatax}%
                  {\@tempswafalse\@citedatax[]}%
}

\def\@citedatax[#1]#2{%
\if@filesw\immediate\write\@auxout{\string\citation{#2}}\fi%
  \def\@citea{}\@cite{\@for\@citeb:=#2\do%
    {\@citea\def\@citea{, }\@ifundefined
       {b@\@citeb}{{\bf ?}%
       \@warning{Citation `\@citeb' on page \thepage \space undefined}}%
{\csname b@\@citeb\endcsname}}}{#1}}%

%
\def\@citex[#1]#2{%
\if@filesw\immediate\write\@auxout{\string\citation{#2}}\fi%
  \def\@citea{}\@cite{\@for\@citeb:=#2\do%
    {\@citea\def\@citea{; }\@ifundefined
       {b@\@citeb}{{\bf ?}%
       \@warning{Citation `\@citeb' on page \thepage \space undefined}}%
{\csname b@\@citeb\endcsname}}}{#1}}%

%
\def\@biblabel#1{}
\makeatother



\newdimen\bibindent
\bibindent=0.0em
\def\thebibliography#1{\section*{\refname}\list
   {}{\settowidth\labelwidth{[#1]}
   \leftmargin\parindent
   \itemindent -\parindent
   \listparindent \itemindent
   \itemsep 0pt
   \parsep 0pt}
   \def\newblock{}
   \sloppy
   \sfcode`\.=1000\relax}